# ***Protect Against Unintentional Insider Threats: The risk of an employee's cyber misconduct on a Social Media Site.***

Guerrino Mazzarolo, Juan Carlos Fernández Casas, Anca Delia Jurcut, Nhien-AnLe-Khac

***Abstract***

Social Media is a cyber-security risk for every business. What do people share on the Internet? Almost everything about oneself is shared: friendship, demographics, family, activities and work-related information. This could become a potential risk in every business if the organization's policies, training and technology fail to properly address these issues.  In many cases, it is the employees' behaviour that can put key company information at danger. Social media has turned into a reconnaissance tool for malicious actors and users accounts are now seen as a goldmine for cyber criminals. Investigation of Social Media is in the embryonic stage and thus, is not yet well understood. This research project aims to collect and analyse open-source data from LinkedIn, discover data leakage and analyse personality types through software as a service (SAAS). The final aim of the study is to understand if there are behavioral factors that can predicting one's attitude toward disclosing sensitive data.

## *I. Introduction*

Until the past decade we have been used to interacting with people face to face. We captured memories with analogic cameras, talked to one another in person, or sent handwritten letters to our family. In a matter of years those everyday common acts have become outdated. The evolution of the internet has brought about a new age of social communication and this phenomenon has extended into every aspect of our modern life. A new model of society is progressing; instant communication, endless engagement, follower counts, superficial engagement, liking, posting, or sharing content are the pillars of a society based on the appearance instead of the being. The number of people using social media has increased significantly, to more than 2.46 billion in 2017 and breaching 3 billion in 2021 (see Figure 1).

*Figure 1. Number of social network users worldwide from 2010 to 2021 (Statista, 2020)*

The influence of social media on businesses, as well as people, cannot be denied. Today, almost every enterprise has its own social media channel enabling businesses to gain exposure, traffic and market insights. However, as with all things, *not all that glitters on social media is gold.*

Social media presents a cyber-security risk for every business. Individuals share almost everything about themselves on the web: friendships, demographics, family, activities and work-related information. This could present a potential risk for businesses if organisational policies, training and technology fail to properly address the issue.  In many cases, it is employees' behaviour that puts key company information in danger. Most personnel lead a very connected life, where they are constantly checking and posting a large amount of information on social media. This can lead some employees to divulge private company data through platforms that they believe are protected and private.

In 2018, the Office of the Information and Privacy Commissioner of Alberta (OIPC) reported that a breach occurred when an employee, who was looking for technical support from a close contact, accidentally sent and revealed spreadsheets containing private information without authorization (OIPC, 2018). People often fail to consider or appreciate how attractive data can be for cybercriminals, state-intelligence gathering, data brokers and marketers. Once any data becomes public, what it is used for is outside of customer's control. The information shared could contribute to a cybersecurity risk and might be difficult to manage or mitigate (Zulkarnaen, Rasdan, Mat Daud, Ghani, Suriatini & Hery, 2016). Social media has turned into a reconnaissance tool for malicious individuals and user-accounts are now seen as a goldmine for cyber criminals.

In fact, any data disclosure could be used for different malicious purpose: phishing and social engineering, intelligence gathering, intellectual property theft, or unfair competition. The problem of cyber-security in relation to social media is real, consistent and continues unabated. In April 2019, the US Federal Bureau of Investigations (FBI) issued a security alert to private sector partners regarding foreign intelligence services using social media accounts to target and engage employees with US government clearance (Cimpanu, 2019).

In 2014, iSIGHT Partners revealed a three-year cyber espionage operation targeting and spying on foreign military and political leaders using social networking. According to the iSIGHT Partners report, hackers used fake accounts on Facebook, Twitter, LinkedIn, Google+, YouTube and Bloggers alleging that they worked in journalism, government, or defence (iSIGHT Partners, 2014). Mika Aaltola, researcher at the Finnish Institute of International Affairs, published a paper focused detailing a Chinese preference for LinkedIn in terms of acquiring classified information from states and enterprises (Aaltola, 2019). These are not unusual examples of intelligence operations nor limited to a unique social media platform. All intelligence agencies use similar activities. When data leakage occurs, security professionals are faced with the same question: ‘how can we prevent this from happening again?’

Over the past decade researchers and practitioners have discussed and examined the causes and characteristics of the perpetrators of insider threats. With the development of risk strategy, it has become clear that mitigation cannot solely rely on security control measures and other security related tools (Mahfuth, 2019). Increasingly research communities have focused their interest on technical and behavioural indicators as well as human factors (Gamachchi & Boztas, 2017).

The investigation misleading information related to insider threats in social media is in an embryonic stage and thus, not well understood. To advance the knowledge in this field there is a continuous need to find new techniques in order to detect and deter insider threats (Holt & Bossler, 2016). The purpose of this paper is to address this challenge and put forward a grounded framework analysing the contributions that have been made to date. Research included examines social media security risks from the unintentional insider view and provides a testing environment based on cybersecurity defence and theories to better understand how human personality engages with this unique domain. A secondary purpose is to verify whether a new indicator from these theories could be developed, with the inherent potential for practical implementation resulting in a reduced overall risk of data breach.

## *II. Insider Threat: A Background*

Insider threats have been present well prior to the existence of technology. For centuries humanity has told stories about infamous attacks coming from trusted people. The quote that can be considered as a mantra for insider threat hunter is: ‘Et tu, Brute?’, a Latin phrase meaning ‘Even you, Brutus?’. It is allegedly attributed to Emperor Caesar at the moment of his assassination in the Senate house, addressed to his beloved nephew Brutus (Shelley, 2013). These words have come to represent an ultimate betrayal from the most unexpected source, such as a trusted partner or family member.

More and more often, cybercrime champions are addressing cases within organisations. The report ‘2020 Cost of Insider Threats: Global’ disclosed that the number of incidents has increased by 47% and the average annual cost of insider threats has also grown by 31% to $11.45 million in the past two years (Ponemon Institute, 2020). In addition, statistic suggest that insider threats account for roughly 30% of all cyber security incidents in government departments and organisations (IBM, 2017). The dangers that come from inside are more difficult to predict and discover because employee are familiar with the organisation’s infrastructure and the security controls applied. This result in accessing effortlessly to classified material (Mazzarolo and Jurcut, 2020). Insider crimes are usually conducted by two types of users: malicious users acting on purpose and employees accidentally causing data breaches and leaks. The result can be the same: data leakage, fraud, theft of confidential information, robbery of intellectual property, and the sabotage of computer systems.

Descriptions of the types of insider threat can be found from different authoritative sources. In this case, the US-CERT explanations are used for the sake of completeness. The ‘Guide to Insider Threats’ defines *intentional* as follows:

> ‘A malicious insider threat is a current or former employee, contractor, or business partner who has or had authorised access to an organisation's network, system, or data and intentionally exceeded or misused that access in a manner that negatively

> affected the confidentiality, integrity, or availability of the organisation's information or information systems' (Capelli, Moore, & Trzeciak, 2012).

The report 'Unintentional Insider Threats: A Foundational Study' defines *unintentional threats* as:

> '… a current or former employee, contractor, or business partner who has or had authorised access to an organisation's network, system, or data and who, through action or inaction without malicious intent, causes harm or substantially increases the probability of future serious harm to the confidentiality, integrity, or availability of the organisation's information or information systems' (CERT Insider Threat Team, 2013, p. 2).

Governmental agencies, security firms and academic researchers have come together to confront the mutual enemy and propose alternative and multidisciplinary solutions (Karampelas, 2017). In order to understand what drives different insiders to illegal deeds, the NATO Cooperative Cyber Defence Centre of Excellence (CCDCOE) established five distinct insider profiles: sabotage, theft (of intellectual property), fraud, espionage, and unintentional insiders (Kont, Pihelgas, Wojtkowiak, Trinberg, & Osula, 2015).

Where sabotage, theft, fraud, and espionage require a deliberate malicious factor the unintentional player probably will not even know they are doing something wrong, but will have inadvertently harmed an organisation's assets through the leaking of data, or providing access to external cybercriminals. Figure 2 shown the type of insider threat and their actions.[1]

*figure 2. insider threat type.*

[1]In the counter-intelligence (CI) field, the acronym MICE (Money, Ideology, Coercion/Commitment and Ego) has been fully accepted by the CI community for decades as the main "motivational and emotional aspects" for the act of disclosing information. Those four factors obviously implied some kind of weakness or vulnerability. Sometimes, a mix of two or three of these factors also are decisive as motivation. Nowadays, an alternative framework is being discussed and accepted by some CI experts. It is the path from MICE to RASCLS, the acronym for *reciprocation, authority, scarcity, commitment (and consistency), liking, and social proof.* According to the former CIA National Clandestine Service (NCS) officer Randy Burkett, today's CI departments often deal with non-state actors with complex mixtures of competing loyalties, including family, tribe, religion, ethnicity, and nationalism. (Burkett, 2013).

Previous research analysed insider threat cases and tried to find common characteristics that lead to an incident. These indicators are essential for cyber security operators to monitor, detect, and response against possible incidents. The common behaviours that could indicate an insider threat can be differentiated between digital and personal behaviours. Some frequent patterns are reviewed by security company Varonis (Petters, 2020). Digital hints are associated with an employee's usage of data, especially if the actions are not directly part of their routine job description. For example, seeking, saving, moving, or printing large amounts of classified information, accessing documents that are not linked to the employee role, or using unauthorised remote storage data. Real life user style could also be a precursor signalling further incidents. Some behaviours such as displaying disgruntled conduct, ethical flexibility, logging in to the firm network during off-hours, repeatedly violating organisational policies or request access exceeding 'need to know' responsibility are all warning signs requiring attention.

Insider threat occurrences can impact organisations in a multitude of aspects however, the riskiest consequences are financial and reputational. In July 2019, the federal court charged engineer Page Thomson with computer fraud and abuse for an intrusion on the stored data' of Capital One (Sheetz, 2019). The banking corporation revealed that it suffered a data breach that exposed hundreds of thousands of customers' personal information. Capital One believed the financial impact of the 2019 breach to be between $100m and $150m because of costs associated with the disaster recovery plan including customer notifications, credit monitoring, technology costs and legal support (Warwick, 2019). Insider attacks can additionally produce loss that is difficult to quantify or recover, such as damage to an organisation's reputation. Cases such as that of Edward

Snowden,[2] Chelsea Manning[3] and Robert Hanssen[4] resulted in gigantic damage to the reputation of the United States government agencies.

In order to reduce the insider threat risk, it is crucial to implement a layered approach including policies, procedures and technical controls. Nowadays, as a countermeasure against data leakage within corporations, the implementation of a comprehensive Insider Threat Programme (InTP) is highly recommended. In deploying this programme, corporations should take into consideration that fact that every organisation has to tailor its approach to meet its unique needs. The Intelligence and National Security Alliance (INSA) provides a framework for implementing an InTP with the Insider Threat roadmap, based on a 13-step model representing those actions that have been taken by current successful programmes in both business and government (INSA, 2015). *Figure 3. Insider Threat Program Roadmap (INSA, 2015).*

### *III. Research problem and current investigative approach.*

---

[2]Edward Snowden (1983) is an American citizen, a former Central Intelligence Agency (CIA) employee and subcontractor (Booz Allen Hamilton Co.) who leaked top-secret information from the National Security Agency (NSA) in 2013. Snowden gradually became disillusioned with the NSA global surveillance programs he was involved since he considered they were a clear intrusion into people's private lives. Although he tried to raise his ethical concerns through internal channels, **nobody paid enough attention to the warnings that Snowden could become (was becoming) an "insider threat". Edward Snowden could be considered as an example of "insider threat" with an ethical commitment**. According to Snowden, he considers himself a *whistle-blower* despite a *leaker* since he did not leak the intel for "personal profit".

[3]Chelsea Manning (1987) born as male Bradley Manning, she got female gender identity in 2013. She was a former US Army soldier working as intelligence analyst posted in Iraq in 2009. She leaked sensitive US intel (up to 750,000 documents) to WikiLeaks. She was imprisoned from 2010 until 2017 when her sentence was commuted. According to several military psychiatrists that assessed Manning's personality and psychology during the trial, Manning had been isolated in the Army while dealing with her gender identity dichotomy. The specialists considered that Manning had the perception that her leaks were positively changing the world. **Chelsea Manning could be considered as an "insider threat" under the parameters of psychological imbalance (gender dichotomy and ego) combined with an ethical commitment for a better world.**

[4]Robert Hanssen (1944) is a former Federal Bureau of Investigation (FBI) senior intelligence officer who spied for the Soviet Main Intelligence Directorate (GRU) from 1979 to 2001. Hanssen sold thousands of classified documents to the KGB for more than $1.4 million in cash and diamonds. The intel provided by Hanssen to the Russians detailed US strategies in nuclear war, military weapons technologies and counterintelligence. He is currently serving 15 consecutive life sentences. According to the US Department of Justice, Hanssen's acts of espionage could be considered "possibly the worst intelligence disaster in US history." **Robert Hanssen is a clear example of "insider threat" with a profit motivation (money).**

While malicious actors within the organisation are the most sophisticated and dangerous to manage, recent investigations highlight that unintentional insider threats represent a major risk for business. These menaces can become also a potential attack vectors for both intentional insiders and external adversaries (Trzeciak, 2017). This study investigates unintentional insider threats (UIT), examining related research and best practice developed up until this point to support a better understanding of its origin. Through the development of a specific use case, based on data disclosure on social media, a preliminary theory is presented regarding potential mitigation strategies and countermeasures. Even if the overall insider threat represents a unique risk for the organisation, UIT denote an exceptional challenge and differ completely from intentional actors in terms of motivation and indicators. Meanwhile, policy, technical controls, monitoring, and incident handling have a decisive impact in detecting, deterring, and responding to malicious threats. The same cannot be said for accidental cases.

The adoption of teleworking, cloud solutions, Bring Your Own Device (BYOD) arrangements and continuous interactions with internet and social media have blurred the separation between work and private life. The result is that organisations need to re-assess their security boundaries in order to implement appropriate protective measure and avoiding leakage of internal information.

The threatscape brings in different scenarios: during their day to day business activities, employees can accidentally click a phishing email, install unapproved software, upload sensitive information to the cloud, or transfer confidential data to unauthorized USBs or via email. Fortunately, technical controls are now available which can block or deter these activities and preserve an organisation's confidentiality and integrity. However, accidental disclosure of sensitive information to social media is often blurred. CERT National Insider Threat Centre acknowledge organisational obstacles to establishing, monitoring, and enforcing policy regarding what personnel publish on social media sites (CERT, 2018).

A large amount of data is regularly self-disclosure by users on Facebook, Instagram, LinkedIn etc. This becomes a real problem when the data disclosed is linked with professional activity such as posting images of the work space, expressing negative views of employers or colleagues and, in the worst cases, sharing classified information publicly.

Despite the impressive scale of information disclosure, very little is understood about what motivates users to disclose personal information. Human error results in the majority of insider threat incidents. Because of the human factor, a multidisciplinary people-centric approach is required as a supplementary tier of defence (Elifoglu, Abel, & Tasseven, 2018).

Supervising employee conduct and maintaining data privacy obligations on social media is a challenge. Enterprises must guarantee employee rights following legal and ethical grounds, whilst ensuring that their online activities do not compromise company reputation or leak classified information. Recommendations for a first deterrence include implementing a social media policy to provide a clear code of conduct and non-disclosure agreements with associated disciplinary procedure in case of employee misconduct or infringement. Certain monitoring tools allow for the tracking of employee comments and negative sentiments regarding businesses (Cross, 2014). Even if those applications help to reduce the risk, they still require tailored configuration and close analysis in order to perform well. Training and security awareness are certainly one of the most effective countermeasures. Proper training influence employees and prevent them from clicking links or prompt them to think twice before posting information on the web (Trend Micro, 2018).

A blend of policy, awareness and technical controls can reduce the number of security incidents resulting from unintentional behaviour by more than 50% (Friedlander, 2016).

Reducing insider threats is not a straightforward task. There are several behavioural indicators that can support investigation and identify where a potential threat is coming from. This, however, should be integrated with trustworthy insider threat detection tools that allow the gathering of full data on user activities. Zuffoletti, CEO of SafeGuard Cyber, advises companies to think about social media both as a vector for threat hunting and part of their attack surface (Sheridan, 2019). Due the scope of the subject, the complexity, and the continuous debate on how to reduce the incident landscape, our research restricted the domain and focused the effort to a) targeting the misuse of employer's information on social media and b) analysis and correlation of human personality based on risk taking tolerance.
Our hypotheses were tested and discussed according to Pareto model.
The Pareto analysis, also called 80/20 rule, undertakes that the large majority of problems (80%) are determined by a little important causes (20%). Pareto analysis framework have been preferred for this work because it allows to discover the most important causes of internal data disclosure in social media based on risky personality and can translate the results obtained in further actionable countermeasures (Powell and Sammut-Bonnici, 2015).
The collecting of evidence regarding employees mishandling information was based on internet search methodology and relied on Social Media Intelligence (SOCMINT). SOCMINT refers to a subset of Open Source Intelligence (OSINT) that collects information exclusively from social media sites. The term 'SOCMINT' was developed by Sir David Omand, Jamie Bartlett and Carl Miller (2012). In this same document the authors stressed that this practice should be restricted to non-intrusive collection from open sources (Omand, Bartlett, Miller, 2012). The Șușnea, Iftene definition of SOCMINT was found to be compliant with this research approach. It defines SOCMINT as a convergence of OSINT designs and web-mining techniques applied to social media information and used to identify as well as understand situations that could become a threat for national security. This model is adapted and moved it to the industry defence strategy (Șușnea, Iftene, 2018). Different tools has been developed for data reconnaissance and intelligence gathering.

Some popular applications for collecting different types of public information are Creepy, Maltego, theHarvester, Recon-ng and many more (Chauhan S. and Panda N. K., 2015). As search engine technique, Nihad at al, recommend the use of Google dork for sophisticated research since it has countless dedicated operators that help advanced and targeted searches (Nihad and Rami, 2018).

Deanna Caputo, behavioural scientist at MITRE's Social, Behavioral, and Linguistic Sciences Department suggested that ‘technology always in some way involves human beings’ and therefore ‘you can't tackle a technological challenge without taking into account human nature’ (Caputo, 2012). Human elements are also a major factor in UIT. Previous research has explored personality traits with the aim of discovering specific characteristic that indicate insider threat. Previous studies stress how personality could put people at risk of cybercrime. Holt et al. , investigated the magnitude to which personality traits and user behaviours involve the likelihood of malicious software infections (Holt, Van Wilsem, Van de Weijer, & Leukfeldt, 2018). Weijer and Leukfeldt (Wijer and Leukfeldt, 2017) examine how the big-five personal attributes can cause exposure to attacks and they found evidence of certain people’s traits directly linked with cyber victimization. Borwell et al. remarked that the human element is recognized as the weakest link in information security, and there is often a connection between behaviour of humans and cybercrimes exploitations (Borwell, Jansen, Stol, 2018).

A particularly interesting study explored insider threat events with malicious intent and proposed a justification across a connection between these and ‘Dark Triad’ personality attributes (difficult personalities with traits as Machiavellianism, narcissism and psychopathy) (Maasberg, Warren, & Beebe, 2015). The above mentioned papers has been used as background for exploring the possible link between unintentional incident and user behaviour. Further analysis has taken into consideration the outcome in previous research works and the correlation on big-five (openness, conscientiousness, agreeableness, extraversion, and neuroticism), DISC (dominance, influence, steadiness, conscientiousness) and cybercrime activity.

Most existing works, however, focus on intentional insider threats with the organisational boundary. Nurse et al. , describe the personality characteristic element as a factor of antecedents or key initial reasons to understanding an individual’s propensity to attack (Nurse, Buckley, Legg, Goldsmith, Creese, Wright, & Whitty, 2014). Additionally, INSA underline that certain personality traits may predispose an employee to acts of espionage, theft, violence, or destruction (INSA, 2017). Personality as a collection of behaviours, cognitions, and emotional patterns has an impact on an individual’s thinking and doing (Cherry, 2019). As a result, this could be of use in terms of indicating possible involvement in activities that could threaten organisations.

Our research was based on an extensive literature review and the assessment of current cyber security defence capabilities to contain UIT. This work permits to identify a specific use case that includes current challenges in an unknown environment, i.e. social media while providing the opportunity to define an innovative countermeasure approach based on personality.

## *IV. Method and data.*

In this section, recommendations include targeted, risk-based approaches that focus on the following two areas: event detection and personality screening. A thorough literature review supported the development of a framework that includes threat vectors and human factors with the aim of discovering a new methodology that is twice as impactful helping to detect data leakage on social media and identify personality traits that can support a preventive cyber defence activity through a process for risk mitigation.

*Figure 4. UIT method framework.*

### Phase 1: Threat vector.

In our research, datasets were collected following the best practice of the social media intelligence discipline (SOCMINT), a branch of open-source intelligence (OSINT) (Schaurer, 2012). SOCMINT describes methods and technologies that allow monitor social media website such as LinkedIn, Instagram, Facebook or Twitter and simultaneously collect publicly available data.

Google dork is a passive information gathering method based on query Google engine against certain specific information. This tool has many special features to help in finding sensitive data that we could apply in this research. More details  about how to use Boolean string to refine and target specific hunt in Google and in order to discover information about companies, employees and geolocations can be found in Johnny Long book ‘Google Hacking for Penetration Testers’ (Long, 2004). Some of the most valuable operators available in Google Dorking are shown in table 1.

*Table 1. Google Dorking Operators*

The variables of interest in our study were: site; country/region; company name; and employee categories (contractor, consultant, full-time, temporary).

The execution of the following query “ site:website.com inurl:in ("Region Area, Country" AND "company name") & ("consultant" OR "contractor" OR "full time" OR "temporary")” resulted in a first list of raw data equal to (n=866) records.

Different methods have been investigated in recent years and various solution has been provided in order to collect data online. Two of these techniques generated great interest and result: web-scraping and application programming interface (API) (Willers, 2017). The final goal of both, web scraping and API, is to retrieve web information. Web scraping extract public accessible data from website beyond the use of software. Differently, API offer direct access and extraction of the data. Our research, through the open source tool Data Scraper, extracted data out of HTML web pages previously settled with Google dork and imports it into Microsoft Excel spreadsheets. The dataset previously obtained was subjected a data wrangling through the application Open Refine including: inspection: detect unexpected, incorrect, and inconsistent data; cleaning: fix or remove any identified anomalies, then verifying: results were inspected to verify correctness.

Following this processes, in our research, n = 470 results with unique and consistent profiles remained.

In the first phase of this analysis, the intention was anomalous detection based on the infringement of security police and non-disclosure agreements on social media. Due to the small number of data points, a manual qualitative analysis was performed (with the support of an automatic string search on python language). This attribute-based analysis took into consideration all sections of LinkedIn, however, the most significant data for investigation were found in the 'Summary and Experience' table. The 'Summary' normally includes name, photo, headline, most recent company, education, contact information and a brief career story. The 'experience section' usually contains job title, company, location, dates of employment and detail information about each job experience. The user detection list was established according to the information disclosure classification in Table 2 and ranked as low or high-risk impact.

*Table 2. Iinformation disclosure classification.*

Social networks promote users to divulge information about their job functions, responsibilities, family, interests, hobbies and beyond, apparently with the scope of engaging them in a wider network with their friends and colleagues. However, the amount and the sensitivity of information disclosed without any form of protection and filter could become extremely risky.

The description details of what you do in your current (or past) job can leverage the disclosure of company classified information. This paper focuses on the leak of internal sensitive information: classified projects, further merger or acquisition, further financial investment in specific area of interest; internal ICT infrastructure: software and hardware in use; sensitive role information: which employees have access to critical systems or key stakeholders, or who has authority within the company; personal information linked with the job: travels, connections, comments.

Qualitative interpretation and evaluation of each profile resulted in a list of n = 120 users that exposed information about employer on social media. Figure 5 shown additional insight.

*Figure 5. User Risk Assessment.*

Fifty-nine percent of users were not relevant and were labelled as low-risk profiles. Forty-one percent of examined profiles demonstrated a certain level of sensitive information disclosed. The sample population of this research is shown in Table 3.

*Table 3. High risk traits*

## Phase 2: Human factor.

The second phase of this research related to human factors and building a personality model for each user. This was a preliminary study constricted by finance, time and manpower. Focus remained on quickly developing and deploying a pragmatic solution. Following careful analysis of open source tools available in the market, several products were tested although two got primary attention for similar characteristics: Crystal and Emma. Both products are based on artificial intelligence and easy to configure. The two tools have been evaluated based on rating and review. Their capabilities were finally verified throughout 10 volunteers that took part in a study aimed to compare, verify and confirm the reliability of the results. Crystal was ultimately chosen because it appeared more mature, accurate and was able to provide with a more in-depth analysis.

Through the use of algorithms that evaluate the communicative content available on LinkedIn and then with statistical modelling, this application judges personalities corresponding to the DISC model classification system (Marston, 2008). DISC is a behaviour assessment focused on four different personality traits: Dominance (D), Influence (I), Steadiness (S), and Conscientiousness (C). Crystal algorithms are assessing the public data available in any profile and provide text-sample analysis from writing style and structure (D'Agostino and Skloot, 2019). Additionally, Crystal handles what others, in your close network circles, have written about you (Figure 6). Crystal can only retrieve information that is being publicly shown. The final goal is to identify people behavioural patterns as accurately as possible. The benefit of using a text-sample approach is the accuracy and the independent framework that avoid third party commitment. The two major flaws are the need of sufficient data sample for the analysis and the possibility of intentionally ingest poisoned data for alter the result. Traits are shown in the Table 4.

*Table 4. DISC personality traits.*

Previous research has demonstrated that personality elements have a degree of impact on internal threats (Xiangyu, Qiuyang & Chandel, 2017). This fact supports the assertion that an evaluation of employee personality traits could be used as indicator in the overall risk screening. Previous literature has also revealed that specific high scores in traits such as extraversion and openness, and low scores in neuroticism, agreeableness, and conscientiousness, correspond with risk-taking behaviour. Specifically, Xiangyu *et al.* emphasised that personality profiles, based on the 'Big Five' methodology (Goldberg, 1990), could be used to predict risk-taking behaviour according to the following description:

> 'high extraversion and openness stream the motivational force for risk taking; low neuroticism and agreeableness supply the insulation against guilt or anxiety about negative consequences, and low conscientiousness makes it easier to cross the cognitive barriers of need for control, deliberation and conformity' (Nicholson, Soane, Fenton-O'Creevy & Willman, 2005).

Research about the influence of personality on self-disclosure of information on social media demonstrates that individuals who are more extroverted disclose more accurate personal information in an attempt to gain more relevance and improve their position on the web (Chen, Pan & Guo, 2016). This correlation supports the Five-factor Model and the DISC personality assessment (Jones, Morris & Hartley, 2013). Following on from this research, as well as the subsequent assessment while reviewing of traits in each model the following parallels to DISC have been made:

- Conscientiousness' = a parallel to DISC personality type C.
- 'Agreeable' = a parallel to DISC personality type S.
- 'Extroverted' = a parallel to DISC personality type I.
- A combination of 'Openness' and 'Neuroticism' = a parallel to DISC personality type D.

In addition, when looking at a DISC profile, both S and C personality styles fall to the more 'introverted' side of the DISC spectrum, while D and I personality styles are considered to be more classically 'extroverted'.

According to the above analyses, the following hypotheses have been made:

- Hypothesis 1: Dominant traits correlate with high-risk taking. These traits include: decisiveness, having a high ego, strength, being a risk taker, and overstepping authority.

- Hypothesis 2: Influence traits correlate with high-risk taking. These traits include: persuasiveness, talkativeness, impulsiveness, being emotional and being more concerned with acceptance than concrete results.
- Hypothesis 3: Steadiness traits correlate with low-risk taking. These traits include: being predictable, understanding, friendly and compliant towards authority
- Hypothesis 4: Conscientiousness traits correlate with low-risk taking. These traits include: sticking to the rules, standards, procedures and protocols.

From the previous phase, 'Threat vector', the users (n=120) was processed by software Crystal and categorized in the four unique DISC groups as shown in Figure 7. The personality traits provide the most accurate possible personality profile based on the information available in LinkedIn (Skloot, 2019).

*Figure 7. Personality Risk Indicators*

**Phase 3: Insider threat prevention.**

These predictions were based on interpretation of previous literature and experiment conducted between March – October 2019 and results have offered significant validation of the proposed hypotheses. The n between threat vector and human factor is summarized in Table 5.

49 incidents have been recorded: 9 associated with dominance traits (18%), 3 with influence (6%), 6 with steadiness (12%) and 31 with conscientiousness (64%).

*Table 5. Data analysis.*

Pareto analysis has been used to correlate both variables, incidents and behavioural characteristics. The evaluation made on the information exposure causes (Figure 8) reveals that the major source to the incidents is related with two types of employee’s traits. 80% of incidents are caused by unintentional actions attributable to conscientiousness and dominant behaviour. Focusing the countermeasure effort in security awareness and additional technical controls to those specific groups could reduce or contain the disclosure of information.

*Figure 8. Pareto Analysis on UIT*

Investigating how personality affects data disclosure on LinkedIn has provided the following findings:

Hypotheses 1, 2 and 3 are positively correlated with the expected data exposure on social media (1 and 2 as high risk-taking profile and 3 as low risk-taking behaviour). Results showed that higher scores on behavioural traits of dominance (H1) and influence (H2) were significantly related to high number of data disclosure. In contrast emotional stability/steady (H3) recorded a low number of disclosure cases. Against research expectation, conscientiousness (H4) shown the highest number of data disclosure. Therefore, H4 was negatively correlated with the results.

- The trait characteristics of dominant personality styles are openness, neuroticism, extroversion and aversion to authority. It was reasonable to hypothesize that staff with traits of openness would be engaged in low levels of data disclosure concerns and as consequences share unnecessary information.

  The influence profile characteristics are extroversion, sociability, and talkativeness; they want to be the centre of attention. I-styles crave interactions with others and, as a result of this, higher disclosure of personal information was expected and confirmed.

- People who demonstrate steadiness personality characteristics demonstrate traits including being: careful, calm, stable, more passive, predictable, and reliable. The lowest number of data disclosure in relation to this group appears to confirm thesis supported in this research.
- Individuals who rate conscientiousness, they strictly follow procedures and standards. They tend to be cautious and contemplative and are not natural risk takers. It was not possible to identify a causal explanation of why this personality trait resulted in the highest in data disclosure.

## *V. Discussion*

This research shows that individual differences in personality can be used as an additional indicator for deterring UIT. Three out of four hypotheses were positively confirmed. Individuals with higher 'dominant' and 'influence' traits were more prone to accepting risk and increase the number of incidents resulting in sensitive data disclosure. On the contrary, individuals with 'reliable' and 'extremely loyal' characteristics were associated with a lower incident rate. A conscientiousness profile could not confirm the hypothesis of low risk-taking. In order to explain that result, an assumption is required based on inductive reasoning. A consciences style, taken to an extreme, could display addiction to work, perfectionism, attention to detail and compulsiveness behaviour traits. Even though conscientiousness positively interacts with psychological well-being, theoretical and empirical work suggest that individuals can be excessively conscientious, resulting in obsessive-compulsiveness, and thereby less positive individual outcomes (Carter, Guan, Maples, Williamson, and Miller, 2015). Their argument does not appear to satisfactorily explain this result however and further analysis is needed. Further analysis should be employed to support an understanding of why this profile was associated with the highest information disclosure.

Limitations and challenges (both theoretical and technical) were encountered during the development of this research. Social media and search engines are continually reviewing their privacy policies. It became difficult for programmers to interact with application program interface (API) and to automate functions that collect and analyse data online. Dataset change occurs continually as users add, modify or remove the data. The ability to interpret or predict what will happen based on behavioural traits is limited. When proposing DISC behavioural styles, it is recommended that consideration is given to style blends, rather than focusing solely on a person's mostly highly scoring trait. Most people will, in fact, show more of some traits and less of others and could possibly have some of all four traits. There are some concerns about the implications that large-scale data mining and analytics could have for society, particularly regarding the impact on privacy, mass surveillance regimes and social bias (Kennedy & Moss, 2015).

## *VI. Conclusion and further work.*

Accidental insiders pose a serious threat to every business. Employees are the most valuable assets of any company, but they can also become the most substantial security threat. Research has indicated that a considerable percentage of cybersecurity incidents and data leakages are caused by a current or former employee acting inadvertently. The expansion of online activity and social networking in recent years has jeopardized security and caused significant losses to organisations due to leakage of information by their own employees (Johnson, 2016).

When assessing the overall risk of unintentional insider threat, it is important to consider different aspects to prevent, detect and respond in case of an incident. The preventative measures currently in use, based on purely technical approaches, are insufficient; defending insider threats requires more than technology (Stahie, 2019). The aim of this research was to develop a framework that provided possibilities for detecting data leakage within an organisation and exploring the personality sphere linked to those incidents, in order to find common characteristics that can support a predictive defence capability. Building a comprehensive IT security program should take into consideration the reduction of blaming and punishment to the end users because often they are the victims. As recommended by CERT, a positive incentive offer the possibility of a more reasonable and beneficial approach to reducing the insider threat with less undesirable consequences (CERT, 2016).

The evolution of technology has progressed to the same extent as security challenges (Schneir, 2018). Threats, arising from internal personnel's activities or lack of awareness, appear to represent a higher risk to information security than challenges triggered by outside attackers (Hekkala, Väyrynen & Wiander, 2012). This study contributed to the perception of how additional protection regarding data leakage from UIT can be achieved on social media. Expanding on previous research, a new framework for insider threat detection and prevention was presented and developed based on social media domains and personality traits. This paper differentiates itself from existing research which is based almost exclusively on technical indicators. The test environment included an attempt to correlate incident detection tracking using advanced techniques, searches, and risk-taking personality traits based on SOCMINT techniques and DISC methodology. Contrary to research expectation, the results obtained shows that conscientiousness traits result  to be the most risky profile in data disclosure. Additionally, steady seam to confirm the aversion of risky attitude and avoid incidents. Once again this proves the complexity of insider threat subject and the difficulties to  encounter trustable indicators that can prevent data leakage.

Even if they have signed non-disclosure agreements and are bound to security policy and regulation, people disclose information. Revealing information about someone to others is part of being human but doing that over social media can be extremely dangerous. A social media audience is unlimited, and cybercriminals are scanning and targeting employees in real-time with the purpose of collecting sensitive data that can be utilised in offensive activity such as social engineering or phishing. The result of information disclosure can have dramatic financial and reputational consequences (Long, Fang & Danfeng, 2017). Incident detection correlated with personality trait analysis could help decrease the overall risk but this needs to be integrated with other indicators. If it is used alone there is a risk it could be over interpreted when it cannot be considered a constant/established trait; people’s actions are influenced by different circumstances. This research was based on risk-taking behaviour however, other relevant human factors can influence daily activity such as fatigue, stress and environmental variables (Carnegie Mellon University, 2013).

Unintentional insider threat presents a problem to security practitioners and academics alike and further research is necessary to develop a more exhaustive understanding of risk tolerance in the context of UIT. In the future, organisations will inevitable rely more on online services such as the Cloud. Cutting edge technology and unpredictable situations (such as the COVID-19 pandemic, where a large portion of the workforce has suddenly transitioned to teleworking) can significantly increase insider risk.

Randy Trzeciak, director of the CERT National Insider Threat Centre, is quoted as having said ‘this extraordinary situation has increased risk factors for insider incident’ (Carnegie Mellon University, 2020). As a result, insider threats will become progressively more complex and difficult to identify. Moving from traditional detection methods to new approaches, such as those analysed in this case study, will soon be insufficient. Technologies such as data science and artificial intelligence might be soon be implemented to support detection of insider threats before they cause irreversible damage (Jou, 2019). In February 2020 during the last RSA conference, one of the most important information security worldwide summits, experts suggested that enterprises should develop their own risk algorithms by combining machine learning capabilities with behavioural analytics (Asokan, 2020). Next steps will concern further work on joint multi-risk domain indicators that could be used as a UIT deterrent.

# FIGURES

**Number of social network users worldwide from 2010 to 2021 (in billions)**

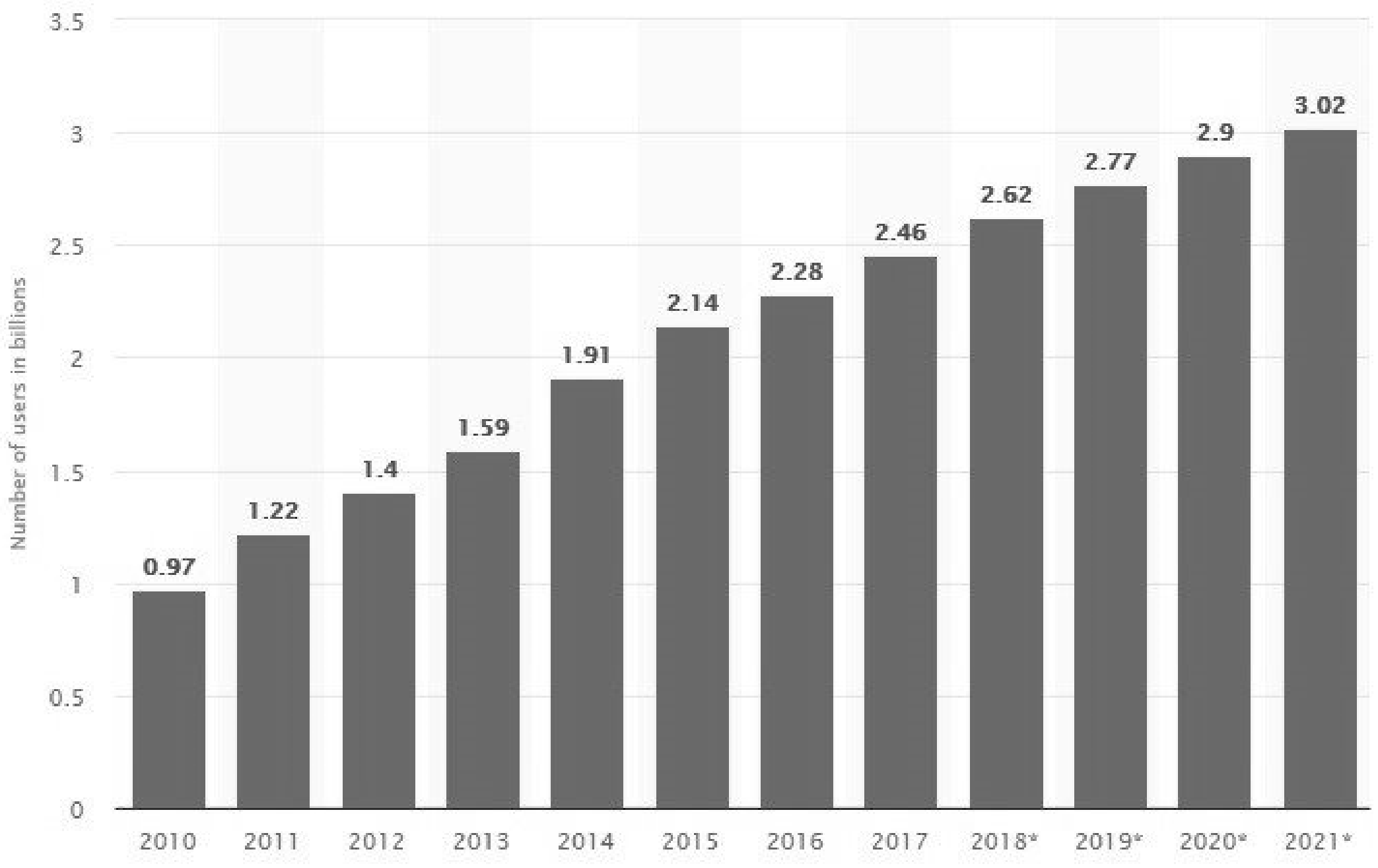

Figure 1. Number of social network users worldwide from 2010 to 2021 (Statista, 2020)

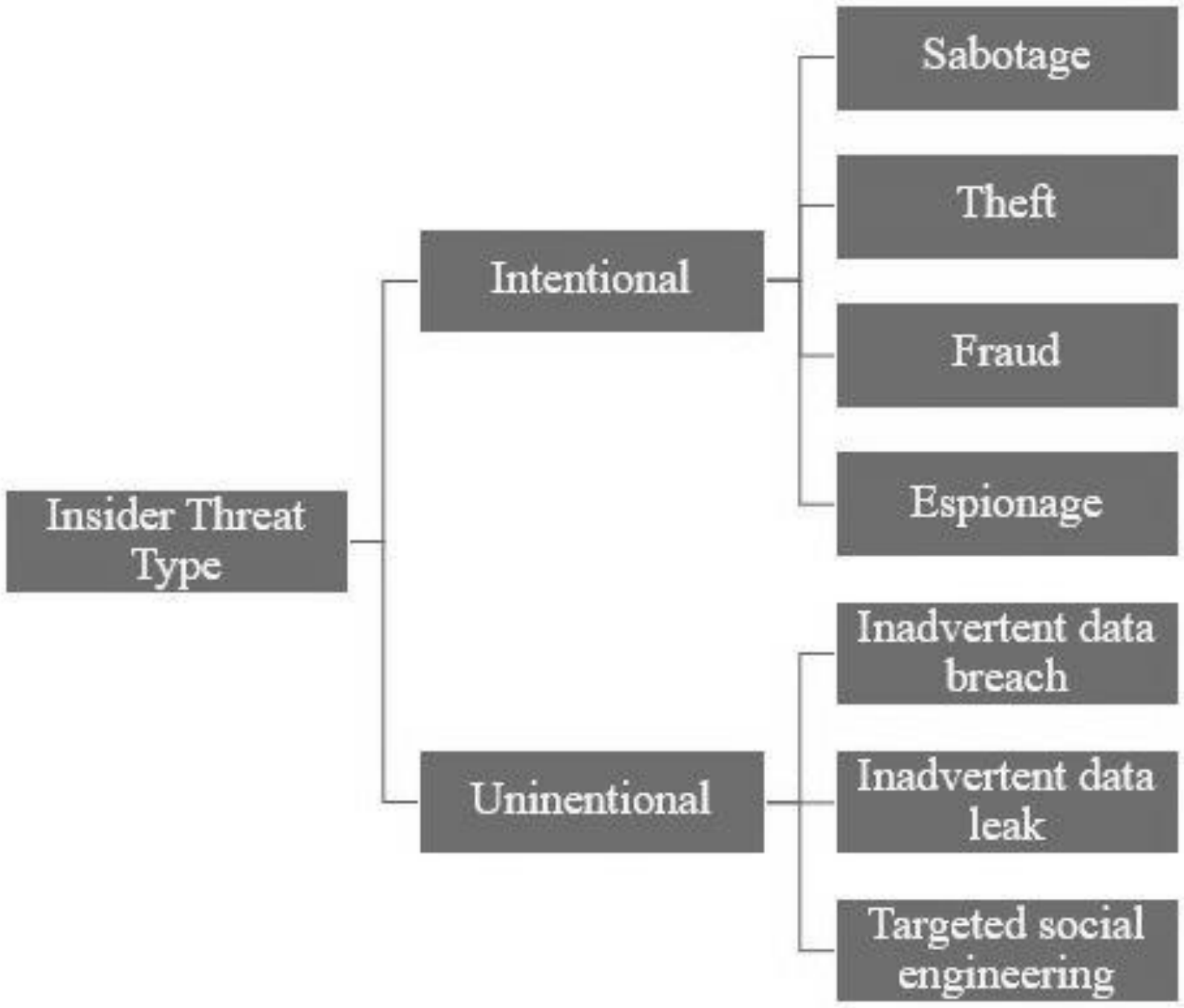

Figure 2. Insider Threat type

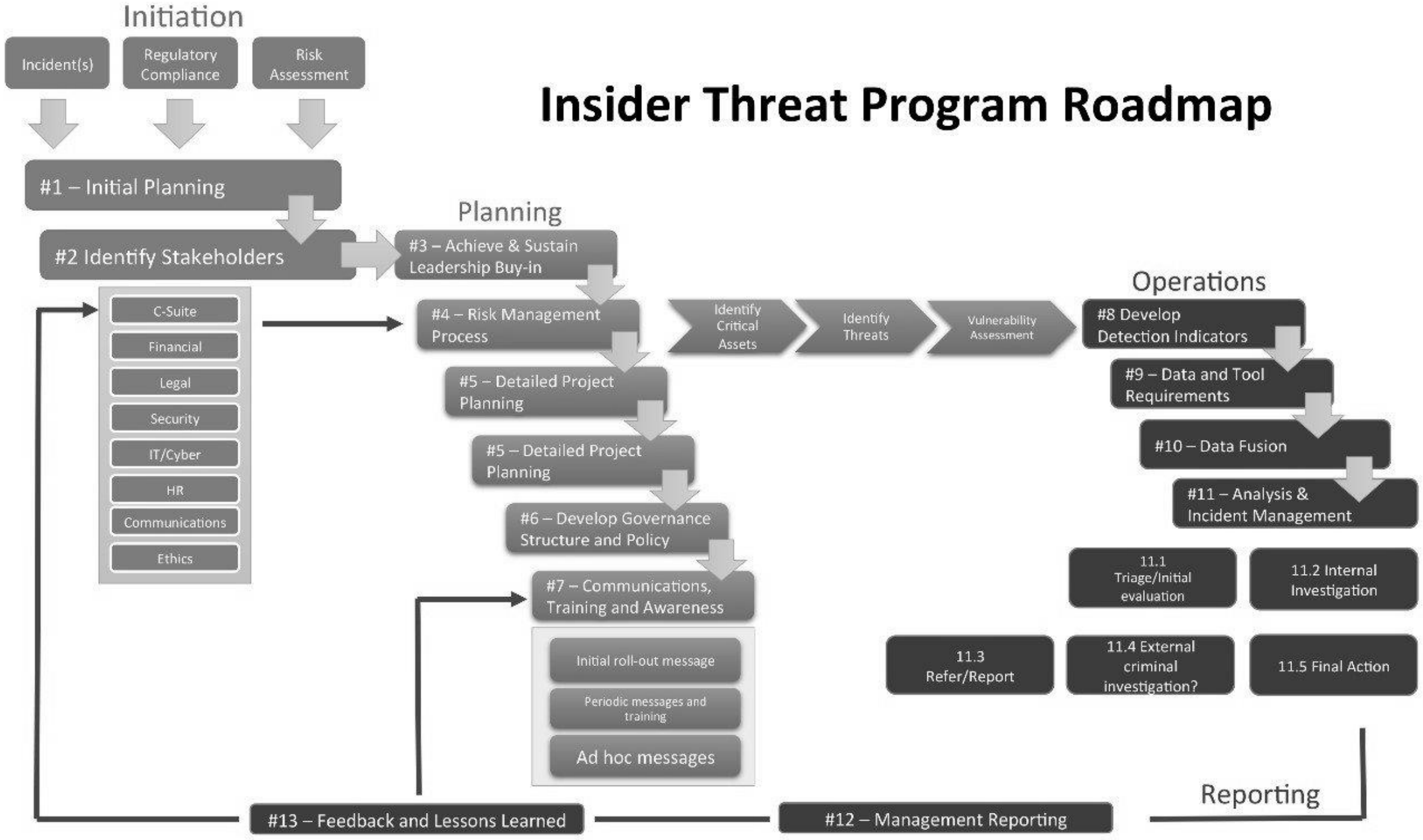

Figure 3. Insider Threat Program Roadmap (INSA, 2015)

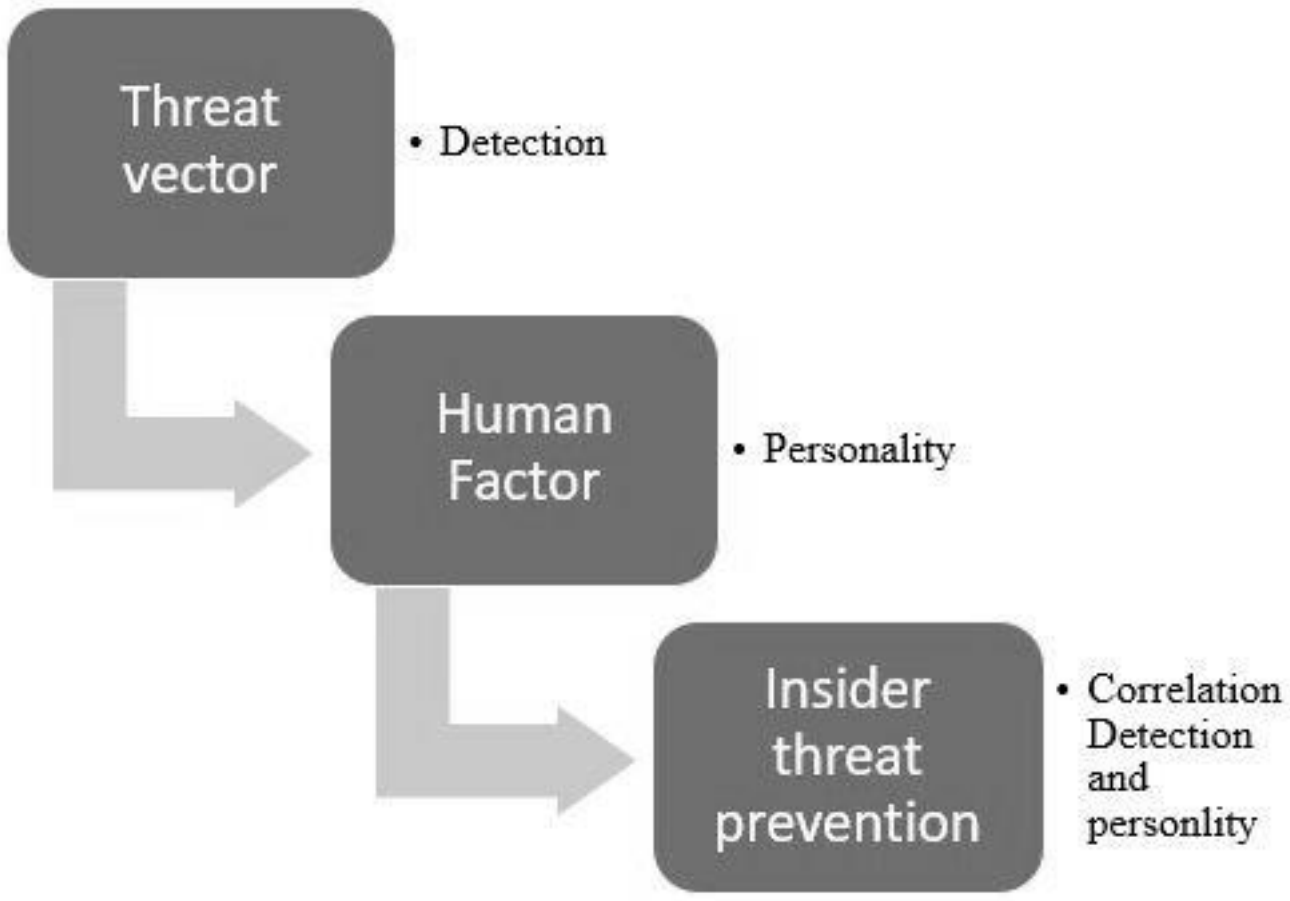

Figure 4. UIT method framework

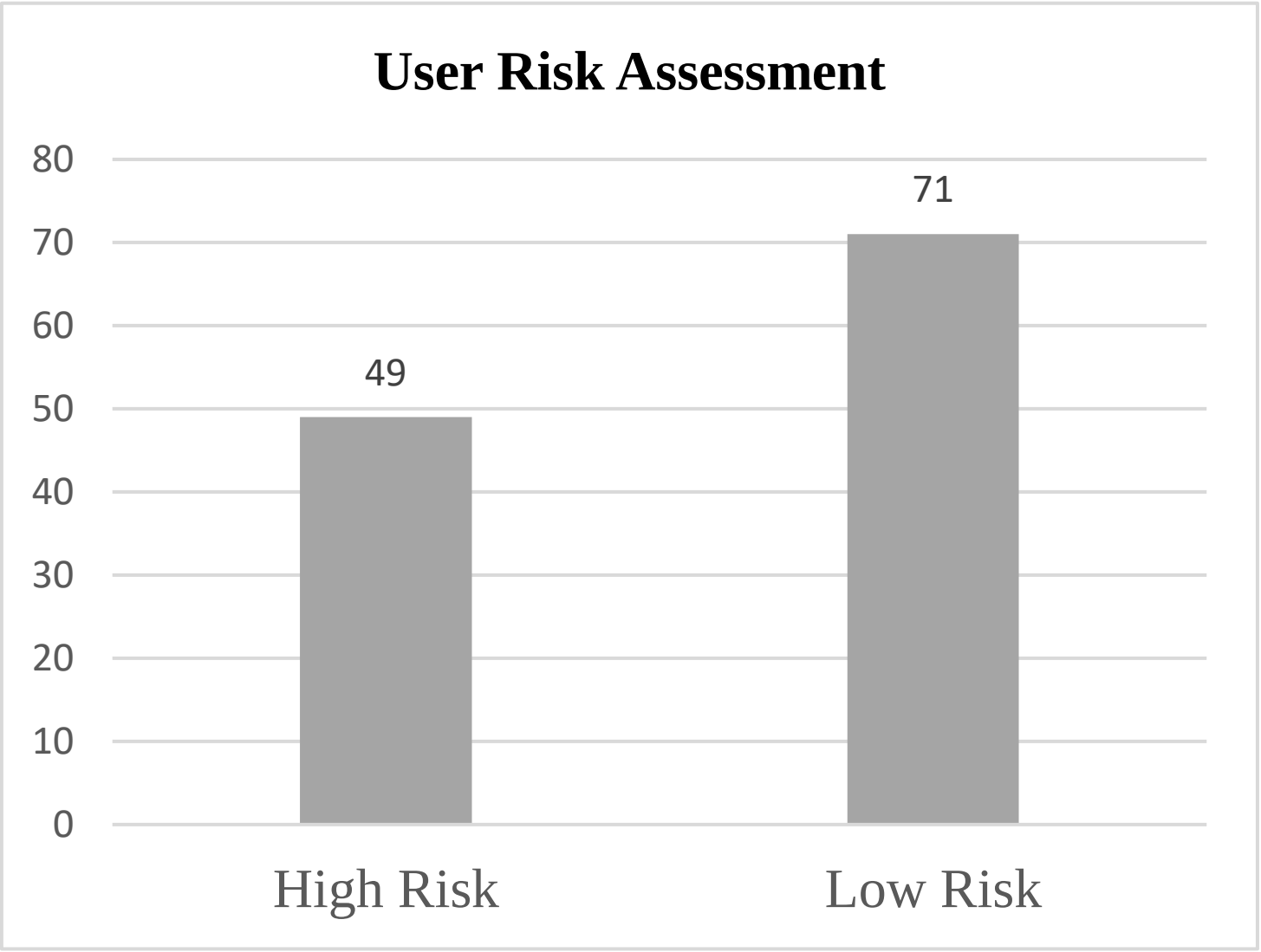

Figure 5. User Risk Assessment

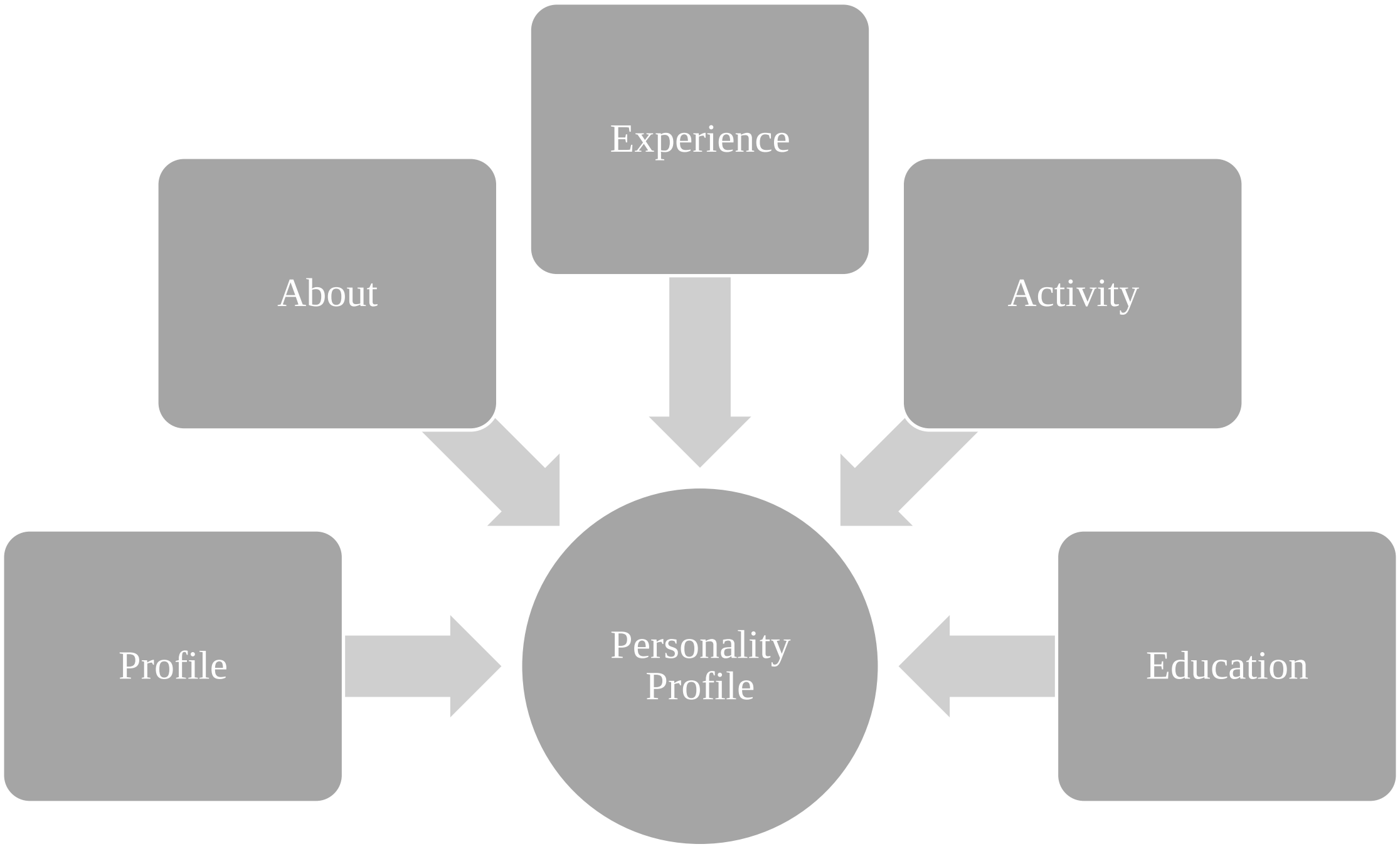

Figure 6. Personality profile.

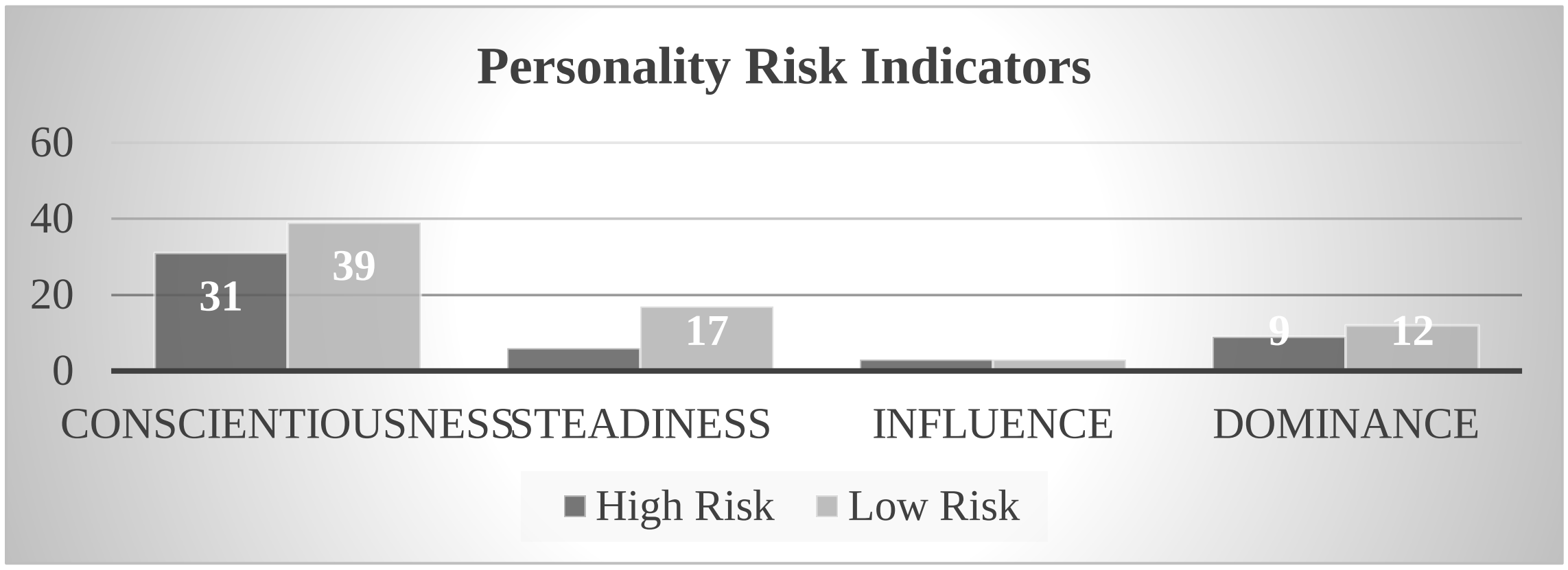

Figure 7. Personality Risk Indicators

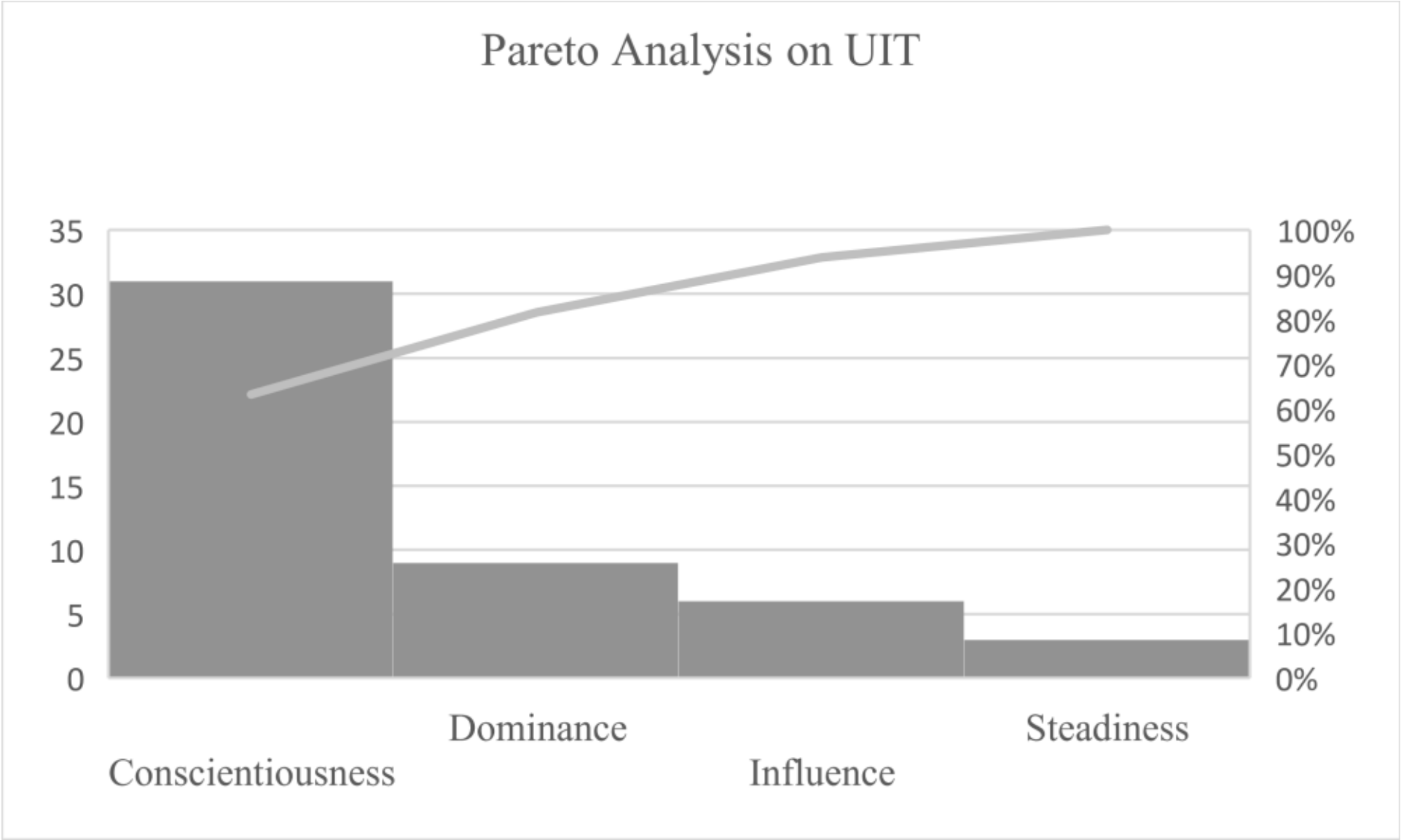

Figure 8. Pareto Analysis on UIT

**TABLES**

*Table 1. Google Dorking Operators*

| Type of information |
| --- |
| Intitle: Looks out for mentioned words in the Page title |
| Inurl: Looks out for mentioned words in the URL. |
| Filetype: This is used to find filetypes. |
| Intext: This helps to search for specific text on the page |
| site: Results from within a specific website |
| OR: Search for one of two keywords |
| AND: Search for one and other keywords |
| "": Search for a specific combination of keywords |

*Table 2. Information disclosure classification*

| Type of information |
| --- |
| Internal sensitive information |
| Internal ICT infrastructure |
| Sensitive role information |
| Personal information linked with the job |

*Table 3. High risk traits*

| Traits | Male<br>n=46 | Female<br>n=3 |
| --- | --- | --- |
| Age | 24-58 years | 24-32 years |
| Work experience | 6 months-18 years | 3 months-3 years |
| Role | 20%Technical-80% Managerial | 100%Managerial |

*Table 4. DISC personality traits.*

| Dominance | Influence | Steadiness | Conscientiousness |
| --- | --- | --- | --- |

| | | | | |
|---|---|---|---|---|
| Pros | Direct<br>Result-oriented<br>Decisive | Inspirational<br>Interactive<br>Outgoing | Patient<br>Tactful<br>Agreeable | Analytical<br>Reserve<br>Precise |
| Cons | Extrovert<br>Neurotics | Extrovert<br>Impulsive | Slow<br>Sensitive | Calculating<br>Condescending |

*Table 5. Data analysis.*

| **DISC** | **Events** | **Incidents** | **%Data Disclosed** |
|---|---|---|---|
| Dominance | 21 | 9 | 18% |
| Influence | 6 | 3 | 6% |
| Steadiness | 23 | 6 | 12% |
| Conscientiousness | 70 | 31 | 63% |